# Neural Dynamics of Delayed Feedback in Robot Teleoperation: Insights from fNIRS Analysis

Tianyu Zhou, Ph.D.[1], Yang Ye[2], Qi Zhu, Ph.D.[3], William Vann, Ph.D.[4], Jing Du, Ph.D.[5*]

[1] Postdoc Research Associate, The Informatics, Cobots and Intelligent Construction (ICIC) Lab, Department of Civil and Coastal Engineering, University of Florida, FL 32611; email: zhoutianyu@ufl.edu

[2] Ph.D. Candidate, The Informatics, Cobots and Intelligent Construction (ICIC) Lab, Department of Civil and Coastal Engineering, University of Florida, FL 32611; email: ye.yang@ufl.edu

[3] Ph.D, National Institute of Standards and Technology, Boulder, CO 80305; email: qi.zhu@nist.gov

[4] Ph.D, The Informatics, Cobots and Intelligent Construction (ICIC) Lab, Department of Civil and Coastal Engineering, University of Florida, FL 32611; email: billyvann@ufl.edu

[5*] Professor, The Informatics, Cobots and Intelligent Construction (ICIC) Lab, Department of Civil and Coastal Engineering, University of Florida, FL 32611; email: eric.du@essie.ufl.edu; Corresponding Author.

**ABSTRACT**

As robot teleoperation increasingly becomes integral in executing tasks in distant, hazardous, or inaccessible environments, the challenge of operational delays remains a significant obstacle. These delays are inherent in signal transmission and processing and can adversely affect the operator's performance, particularly in tasks requiring precision and timeliness. While current research has made strides in mitigating these delays through advanced control strategies and training methods, a crucial gap persists in understanding the neurofunctional impacts of these delays and the efficacy of countermeasures from a cognitive perspective. Our study narrows this gap by leveraging functional Near-Infrared Spectroscopy (fNIRS) to examine the neurofunctional implications of simulated haptic feedback on cognitive activity and motor coordination under delayed conditions. In a human-subject experiment (N=41), we manipulated sensory feedback to observe its influences on various brain regions of interest (ROIs) response during teleoperation tasks. The fNIRS data provided a detailed assessment of cerebral activity, particularly in ROIs implicated in time perception and the execution of precise movements. Our results reveal that certain conditions, which provided immediate simulated haptic feedback, significantly optimized neural functions related to time perception and motor coordination, and improved motor performance. These findings provide empirical evidence about the neurofunctional basis of the enhanced motor performance with simulated synthetic force feedback in the presence of teleoperation delays.

**INTRODUCTION**

Robot teleoperation enables human operators to command and control robots in distant, hazardous, or inaccessible environments (Senft et al. 2021). This ability expands the range of feasible applications, such as deep-sea exploration, space missions, and hazardous material handling, allowing for complex tasks to be conducted beyond the conventional spatial limitations imposed between the human operator and the robot (Zhou et al. 2023). Despite its remarkable potential, robot teleoperation is persistently confronted with the inherent challenge of operational delays (Wenhao et al. 2017). These delays present as a significant barrier, creating latency in the transmission of commands and execution of robotic actions, which are the result of intrinsic physical constraints such as signal transmission distances and computational processing boundaries (Kluge et al. 2013; Payra et al. 2020). Consequently, these delays become critical obstacles that adversely affect the operator's situational awareness, precision in control, and overall task performance (Orlosky et al. 2018). This consequences also include increased cognitive workload, heightened potential for errors, and challenges in maintaining operational efficiency and effectiveness in teleoperation scenarios (Kim et al. 2021).

In order to mitigate the implications of inevitable delays in robot teleoperation, literature has presented a variety of technical or behavioral countermeasures (Farajiparvar et al. 2020).

Prominent among these countermeasures include supervisory controls (Manoharan and Ponraj 2019), predictive controls (Uddin and Ryu 2016), diversified of interaction modalities (Magrini et al. 2020), and intensive trainings for developing adaptive manipulative tactics such as the 'move and wait' strategy (Hokayem and Spong 2006). These countermeasures aim at optimizing the reactive actions based on the predicted delay patterns (Zhu et al. 2023), or improve human responses while repetitive training (Pervez et al. 2019). Nevertheless, these existing methods are less effective when delay patterns are less clear, or when training is limited such as in emergent scenarios. To prepare for more extreme conditions of delayed teleoperation, we have proposed an innovative approach to sensory manipulation. By utilizing a physics engine, we simulate synthetic force feedback in anticipation of the actual haptic signal data (Du et al. 2023). This method creates a more intuitive and responsive teleoperation experience, even when communication delays change. The simulated feedback is designed to approximate the real physical interactions that the robot would experience, providing the operator with a preemptive sense of the forces involved in the task. In our pilot test we have found that this sensory manipulation method could significantly improve the operator's perception and control, thereby reducing the adverse effects of the inherent delays in robot teleoperation.

However, we noticed a knowledge gap in terms of the neurofunctional underpinnings of sensory manipulation or other similar approaches as countermeasures to teleportation delays. While existing studies have examined the implications of teleoperation delays and corresponding mitigation strategies on motor performance, or self-assessment of perception and cognitive status, there remains a significant gap in understanding how these strategies affect neural functions, particularly those related to time perception and motor coordination. The existing literature largely neglects the neural underpinnings that could play a crucial role in determining the efficacy of teleoperated manipulations. Specifically, there is a scarcity of evidence on how synthetic, simulated haptic feedback influences these neural processes. This omission is critical as understanding the neurofunctional impacts of sensory manipulation could provide deeper insights into the mechanisms through which these strategies improve teleoperation performance.

The objective of this paper is to address this knowledge gap by exploring the neurofunctional implications of synthetic haptic feedback in delayed robot teleoperation. To this end, we have conducted a human-subject experiment (N=41), utilizing functional Near-Infrared Spectroscopy (fNIRS) to monitor neural activity. This study aims to provide empirical evidence on how simulated force feedback influences neural functions related to time perception and motor coordination, thereby offering a neuroscientific perspective on the effectiveness of sensory manipulation in enhancing teleoperated task performance. Our findings are anticipated to contribute substantially to the field of teleoperation, offering novel insights into the neural dynamics underpinning human-robot interaction in the context of latency challenges. The remainder of the paper introduces the relevant body of literature, the design of the experiment, and the key findings.

## LITERATURE REVIEW

### *Neural Functions in Temporal Motor Tasks in Teleoperation*

Understanding the interplay between neural functions, specifically time perception and motor coordination, in the context of robot teleoperation is essential, especially when considering the challenges imposed by teleoperation delays. This literature review delves into the significance of these functions and how they are impacted by teleoperation latency. The first noticeable function is the time perception ability. The role of time perception in tasks requiring precise timing, such as in surgical procedures or precision engineering, is critical. Studies like Block and Zakay (1996) have explored the subjective nature of time perception, indicating its susceptibility to various factors, including task complexity and attentional resources. Ivry and Spencer (2004) further emphasize the intrinsic link between time perception and motor functions, particularly in tasks requiring synchronization and rhythm. In teleoperation, where delays are common, this relationship becomes even more crucial. Research by Merchant et al. (2013) has demonstrated how altered time perception due to latency can impact the synchronization and timing of motor responses, leading to potential inaccuracies in teleoperated tasks. Literature has attempted to identify neurofunctional activities indicative of these key functions related to temporal motor tasks such as teleoperation. It has been found that the basal ganglia play a central role in timing and time perception, particularly in the milliseconds to seconds range crucial for teleoperation tasks (Merchant et al. 2013). Additionally, the supplementary motor area (SMA) and pre-SMA are involved in integrating temporal and motor information, essential for planning and timing movements (Halsband et al. 1993). Furthermore, the dorsolateral prefrontal cortex (DLPFC) is implicated in the cognitive aspects of time perception (Wei-Cong et al. 2015). Studies by Yin et al. (2019) and Onoe et al. (2001) suggest the DLPFC's role in temporal discrimination and the cognitive control of time estimation, crucial for adjusting to delays in teleoperation. In the context of teleoperation, where operators need to integrate temporal judgments with motor coordination and decision-making, the role of the DLPFC could be significant. It may contribute to how operators perceive and adjust to delays, particularly in tasks that require them to maintain and manipulate temporal information over short periods.

Similarly, motor coordination is vital for executing complex tasks through teleoperated systems, heavily influenced by the quality and timeliness of sensory feedback. Studies by Ankarali et al. (2014) on sensory feedback in motor control underscore the importance of timely and accurate haptic feedback for effective motor coordination. Further, research by Tin and Poon (2005) on internal models in sensorimotor integration suggests that delays in feedback can disrupt these internal models, leading to a misalignment between intended and executed actions. The impact of this misalignment in high-precision tasks, as highlighted in the work of Jones and Kandathil (2018), underscores the necessity for real-time or predictive sensory inputs in teleoperation. Literature has provided solid evidence about the neurofunctional ROIs related to the motor coordination. For example, the primary motor cortex, as shown by Hari et al. (1998), is pivotal not only in movement execution but also in motor planning, adapting strategies in dynamic environments typical of teleoperation. Scott (2012) and Albert and Shadmehr (2016) further illustrate its role in encoding movement parameters and adapting motor plans in response to feedback, crucial under teleoperation delays. Complementing this, the cerebellum, highlighted in studies by Fautrelle et al. (2011) and Johnson et al. (2019), plays an essential role in fine-tuning movements and error correction, ensuring smooth and coordinated motor output. Its involvement in predictive motor control, as noted by Witney et al. (1999), is particularly relevant for anticipating and compensating for communication delays in teleoperation. The synergy between the primary motor cortex and the cerebellum, as discussed by Galea et al. (2011), is fundamental in maintaining precision and control, adapting, and compensating for the delayed feedback inherent in teleoperated tasks.

It is also noted that investigating how simulated feedback influences specific brain regions can provide critical insights into the neural mechanisms that could mitigate the adverse effects of teleoperation delays. The concept of predictive coding suggests that the brain is not a passive recipient of sensory signals but actively generates predictions about incoming sensory information, updating these predictions as new data arrives (Kilner et al. 2007). This model has profound implications for understanding how simulated feedback might be integrated into neural processes to counteract the disorienting effects of delayed teleoperation. Research by Shadmehr et al. (2010) builds on the predictive coding framework, proposing that the brain's predictive mechanisms allow for smoother motor control by anticipating sensory events. This is particularly relevant when considering the DLPFC and its role in cognitive functions, including the integration of sensory information with motor planning (Abe and Hanakawa 2009). Simulated feedback, when designed effectively, could harness these predictive mechanisms, potentially reducing the cognitive load and improving motor execution in teleoperation scenarios. The SMA and pre-SMA, regions involved in the initiation and temporal organization of movements (Shima and Tanji 1998), may also benefit from simulated feedback. By providing early sensory cues, simulated feedback could help in 'pre-setting' these regions, allowing for more accurate timing predictions and motor responses despite delays (Kilavik et al. 2014). This study mainly relies on fNIRS data for capturing the key neurofunctional characteristics, which will be introduced in the next section.

*fNIRS Methods in Exploring Neurodynamic in Teleoperation*

Functional Near-Infrared Spectroscopy (fNIRS) utilizes near-infrared light to monitor brain activity. It operates on the principle that oxygenated and deoxygenated hemoglobin in the brain have distinct absorption spectra in the near-infrared range. When neurons are active, they consume more oxygen, altering the balance between oxygenated and deoxygenated hemoglobin (Zimeo Morais et al. 2018). fNIRS detects these changes, providing an indirect measure of neural activity. This method is advantageous for its non-invasiveness, portability, and relative insensitivity to motion artifacts compared to other neuroimaging techniques, making it suitable for use in diverse settings, including those that simulate real-world teleoperation environments (Tak and Ye 2014).

Compared to other neuroimaging tools like functional Magnetic Resonance Imaging (fMRI), Electroencephalography (EEG), and Positron Emission Tomography (PET), fNIRS offers unique advantages in the context of teleoperation studies (Abtahi et al. 2020). fMRI, while offering high spatial resolution, is limited by its need for a highly controlled, immobile environment, making it less suitable for dynamic tasks (Ma et al. 2022). EEG, with its excellent temporal resolution, is sensitive to electrical noise and requires complex setups (Parvizi and Kastner 2018). PET, though powerful in metabolic studies, involves exposure to radioactive tracers, limiting its practicality (Slough et al. 2016). In contrast, fNIRS is more adaptable to naturalistic settings, relatively motion-tolerant, and does not require a strictly controlled environment. This makes fNIRS a more feasible option for teleoperation research compared to these other methods (Balardin et al. 2017). Furthermore, when compared to subjective self-report measures like the NASA Task Load Index (NASA-TLX) (Hart and Staveland 1988), fNIRS provides a more direct, objective measure of brain activity. While questionnaires can capture an operator's self-perceived workload and stress, they are limited by subjective biases and post-task rationalization. fNIRS, on the other hand, allows for the investigation of real-time neural processes underlying task performance (Maior et al. 2014).

    fNIRS has been instrumental in revealing various neural functions critical in teleoperation tasks. For instance, it can assess the activation of the prefrontal cortex, a region associated with executive functions, decision-making, and adapting to new or challenging situations, all essential in managing the complexities of teleoperation (Euston et al. 2012). Furthermore, fNIRS can explore the activity in motor areas of the brain, like the primary motor cortex, which is directly involved in executing movement commands (Sanes and Donoghue 2000). This is particularly relevant in understanding how operators physically interact with teleoperated systems and adjust their motor responses in real-time. Additionally, fNIRS can be used to examine regions associated with sensory integration and processing, providing insights into how operators combine visual,

auditory, and haptic information during teleoperation (Zheng et al. 2023). fNIRS also plays a pivotal role in providing direct insights into the neural mechanisms underpinning operator's motor performance. In the dynamic and often demanding context of teleoperation, where operators must continually adapt to feedback delays and complex control tasks, fNIRS offers a unique window into the cerebral processes involved (Zhu et al. 2021). By tracking brain activity in real-time, fNIRS allows researchers to observe how different teleoperation conditions, such as variations in feedback delay, affect specific brain regions. This insight is especially valuable in identifying which aspects of teleoperation are most cognitively demanding and how different sensory manipulations can mitigate these challenges.

The ability of fNIRS to detect changes in brain activation patterns provides crucial information for the design and optimization of teleoperation systems. For instance, understanding how operators' brain activity varies with different feedback modalities can guide the development of interfaces that are more aligned with human cognitive processes, thereby enhancing efficiency and reducing cognitive strain (Benitez-Andonegui et al. 2020). Furthermore, fNIRS data can inform the development of training programs that more effectively prepare operators for the cognitive demands of teleoperation tasks, ultimately leading to improved performance and a reduction in operational errors (Naseer and Hong 2015). Thus, fNIRS emerges not only as a tool for scientific inquiry but also as a practical instrument for refining teleoperation technology to better suit the cognitive profiles of its users.

**FNIRS-BASED ANALYTICAL PIPELINE**

*fNIRS System*

In our exploration of the cognitive impacts of delay in teleoperation, we utilized the NIRx fNIRS device, a state-of-the-art neuroimaging tool known for its precision in capturing cerebral hemodynamic responses. Our NIRx device's setup included 16 sources and 15 detectors, with an additional detector at the right pre-auricular (RPA) point acting as a reference to filter out extracerebral signals and systemic interferences. The device operated at a sample rate of 10 Hz, which is standard for capturing dynamic cerebral responses during active tasks. Each source emitted two distinct near-infrared wavelengths, typically around 760 nm and 850 nm, allowing for the differentiation between oxygenated and deoxygenated hemoglobin.

Our study concentrated on analyzing data from several key brain regions relevant to robot teleoperation: the anterior prefrontal cortex (APFC), left and right dorsolateral prefrontal cortex (LDLPFC and RDLPFC), left and right premotor cortex (LM1 and RM1), and left and right primary motor cortex (LPM and RPM) as illustrated in **Fig.1**. Each selected region plays a crucial role in teleoperation: the APFC is involved in executive functions and complex problem-solving (Euston et al. 2012), the LDLPFC and RDLPFC in working memory and decision-making processes (Philiastides et al. 2011), the LM1 and RM1 in movement planning (Hoshi and Tanji 2000), and the LPM and RPM are involved in the execution of movements (Schnitzler et al. 1997). Notably, the APFC, LDLPFC, and RDLPFC also contribute to the perception of time, a cognitive function that becomes especially important in the context of feedback delays where the brain must reconcile the discrepancy between expected and actual sensory inputs (Wei-Cong et al. 2015).

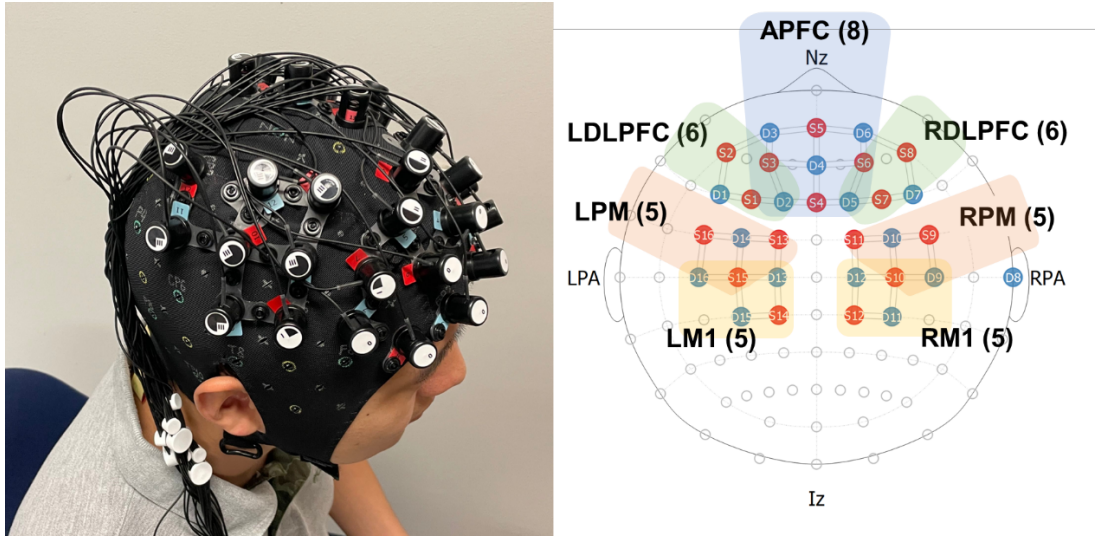

**Fig.1** fNIRS layout setting. (a) Wearing of real fNIRS; (b) Region of Interest (ROI)

To ensure the integrity of our fNIRS data, a thorough cleaning process was implemented to remove physiological interferences, such as those caused by heartbeats and respiration (Pinti et al. 2020). Importantly, the hemodynamic response measured by fNIRS often exhibits a time lag between stimulus presentation and peak response, typically ranging from 2 to 8 seconds (Huppert et al. 2006). This time-to-peak aspect was carefully considered in our analysis, acknowledging the delay inherent in the brain's hemodynamic response to external stimuli, which is critical for accurately interpreting neural activity in the context of teleoperated task execution.

*fNIRS Data Analysis*

The processing of fNIRS data began with MNE-python, a library that can remove secondary noise from the raw data and convert the intensity time series into concentration changes of oxygenated hemoglobin and deoxygenated hemoglobin ($\Delta HbR$) (Gramfort et al. 2013). The pipeline for fNIRS data analysis is illustrated in **Fig.2**. Upon importing the raw fNIRS data, it was converted into optical density ($\Delta OD$), a measure reflecting changes in light absorption due to variations in chromophore concentration in the brain tissue (Tak and Ye 2014).

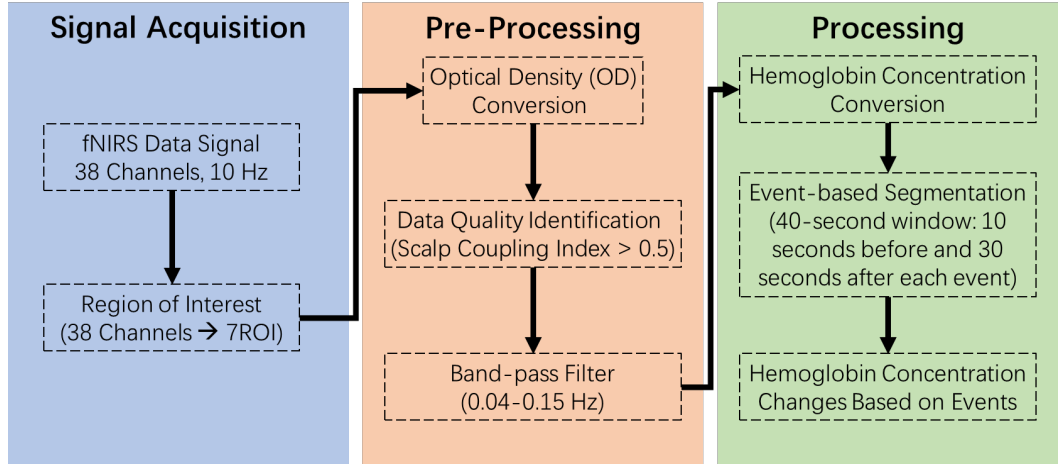

**Fig.2** Analytical pipeline for fNIRS data

An essential step in ensuring data quality involved the evaluation of the Scalp Coupling Index (SCI), an objective metric quantifying the quality of the optode-scalp connection (Pollonini et al. 2016). The SCI is critical in fNIRS data analysis as it reflects the signal strength and integrity; values less than 0.5 typically indicate poor data quality, possibly due to motion artifacts or insufficient contact between the optodes and the scalp. Channels with SCI values below this threshold were excluded from subsequent analysis to maintain the integrity of our dataset.

Following the quality assessment, the optical density data from the fNIRS device underwent a critical filtering process to isolate the neural signals related to cognitive activity from extraneous physiological noise. We employed a finite impulse response (FIR) method, utilizing a bandpass filter within the frequency range of 0.04 Hz–0.15 Hz to target various types of noise (Khan et al. 2020; Pinti et al. 2020): *Cardiac Cycles*: Typically, cardiac-related fluctuations occur at frequencies around 1.0 to 1.5 Hz; *Respiration*: Respiratory patterns generally manifest in the fNIRS signal at frequencies around 0.3 Hz; *Very Low-Frequency Drifts*: Low-frequency drifts in fNIRS data, typically below 0.01 Hz, can arise from slow shifts in sensor positioning or gradual changes in baseline physiological states. The transition band width was set to 0.1 Hz and 0.02 Hz at the high and low cut-off frequencies to ensure a smooth transition between the passband and the stopband, preventing the abrupt cutoff of relevant signals. The high cut-off frequency was designed to exclude high-frequency noise, such as electronic interference or rapid motion artifacts, while the low cut-off frequency was adjusted to remove the slower physiological oscillations without affecting the integrity of the cognitive-related hemodynamic signals.

To obtain measurements of hemoglobin concentration changes, the filtered optical density signals were then converted using the Beer-Lambert Law, which describes the absorption of light in a medium (Swinehart 1962). The law states that the concentration of a light-absorbing substance is directly proportional to the absorption of light as it travels through a given path length. In the context of fNIRS, this principle allows us to estimate changes in oxygenated (HbO) and deoxygenated (Hb) hemoglobin concentrations based on the absorption properties of blood, using a partial pathlength factor to account for the scattering of light in biological tissue. In our analysis, we focused on HbO as the primary measure due to its higher sensitivity to changes in cerebral blood flow, particularly in tasks involving motor execution (Pereira et al. 2023). This decision was supported by literature suggesting HbO's superior reliability in reflecting the brain's response to motor-related demands.

The analytical approach based on specific events, object pick-up and drop-down, which is critical in the teleoperation task. We segmented the data into epochs extending from 10 seconds before to 30 seconds after each event, creating a 40-second window to capture the hemodynamic responses. This response typically exhibits an inherent delay, known as the time-to-peak, ranging from 2 to 8 seconds between the onset of a stimulus and the peak of neural activity. However, the hemodynamic response doesn't immediately return to baseline after reaching its peak; instead, it gradually decreases over several seconds (Amiri et al. 2014; Duarte et al. 2023). Therefore, the entire process, from the initial rise in response to its peak and subsequent decline to baseline, can extend well beyond the time-to-peak, a sufficiently large window is required to capture these dynamics. Choosing this specific time frame for our event-related analysis over a full-trial average approach allows us to focus on the nuanced changes in brain activity that are directly related to task performance. Averaging the data across the entire trial could potentially obscure these detailed event-specific hemodynamic patterns, especially considering the relatively longer periods of lower neural activity between these critical task events.

## MATERIALS AND METHODS

### Overview

The study was approved by the Institutional Review Board (IRB) of the University of Florida, Gainesville, FL, USA (No. IRB202100257). Written informed consents were obtained from all participants in full accordance with the ethical principles of the relevant IRB guidelines and regulations. All methods were carried out in accordance with relevant guidelines and regulations. The following inclusion criteria were applied: (1) age≥18 years; (2) no known physical or mental disabilities; (3) no known musculoskeletal disorders.

### Experiment Task

The main task in the human-subject experiment was an object manipulation task. Participants needed to interact with four colored cubes: grey, green, blue, and purple. Each cube aligned with a target with the same color, requiring participants to accurately move these cubes following a predefined sequence: grey, green, blue, and then purple. The sequenced tasks were systematically structured to gradually increase in complexity and challenge. In this setting, each cube's path to its corresponding target was blocked by various obstacles, which carefully integrated into the task environment. These obstacles vary in size and position, adding to the complexity of the task and representing different locomotor challenges that participants had to contend with as illustrated in **Fig.3**.

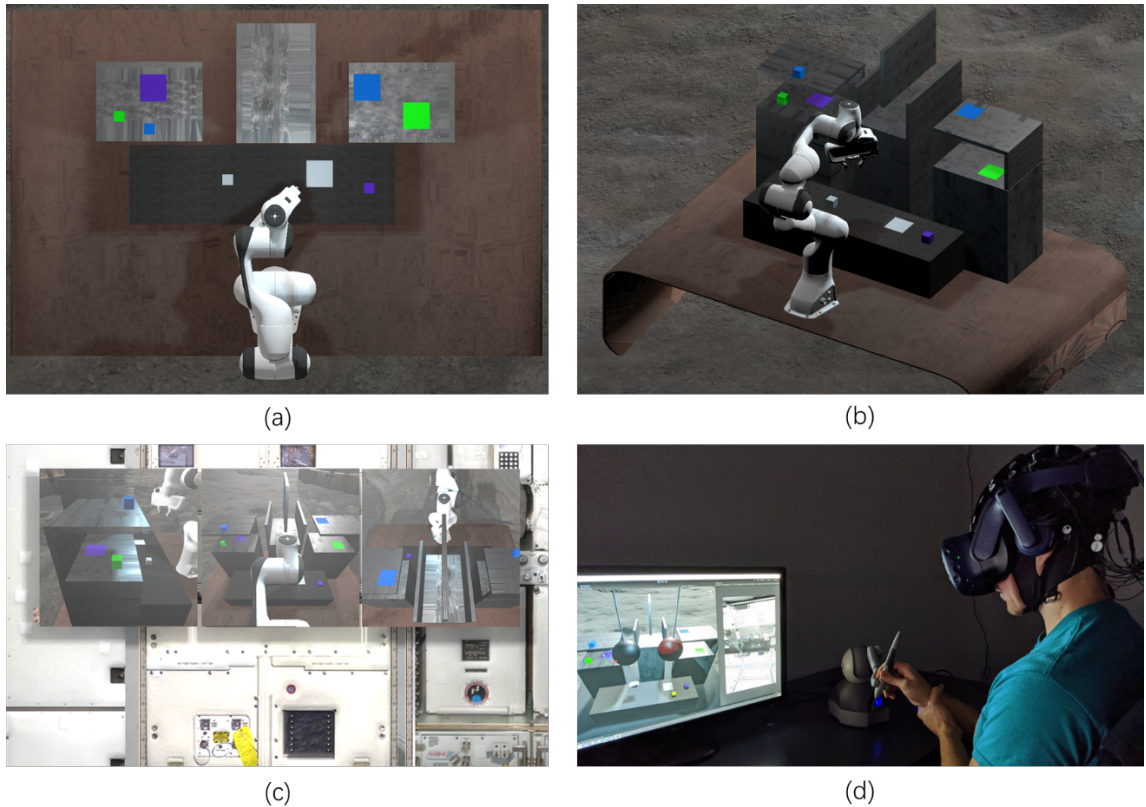

**Fig.3** The layout of the object manipulation task in human-subject experiments. (a) Top view of experimental scene; (b) 3D perspective of the experimental scene; (c) First person view of the participants; (d) Participants completed the experiment using haptic feedback device

When dissecting the delay, it is categorized into haptic feedback delay ($\Delta_{haptic}$), and visual feedback delay ($\Delta_{visual}$). As illustrated in **Table 1**, our experiment was based on four sensory manipulation conditions as follows:

*Condition 1: Control condition*: $\Delta_{haptic} = \Delta_{visual}$, resulting in instantaneous haptic and visual feedback. In this real-time interaction scenario, the operator receives immediate multisensory feedback post-action initiation.

*Condition 2: Anchoring*: $\Delta_{haptic} = 0$ while $\Delta_{visual}$ changes. Due to the intrinsic delays in visual feedback, real-time haptic responses are generated post-action initiation based on the simulated force feedback (e.g., inertia, resistance, vibration) at the local workstation.

*Condition 3: Synchronous*. Both $\Delta_{haptic}$ and $\Delta_{visual}$ are intentionally subjected to a synchronized delay in order to promote multisensory alignment and enhance the coherence of perceptual experiences through the alignment of sensory modalities.

*Condition 4: Asynchronous*. This condition embodies variable delays in sensory feedbacks, presenting a realistic and challenging scenario in which perceptible delays influence the initiation and reception of haptic and visual feedbacks.

**Table 1**. Feedback delays correspond to each condition.

| Condition | Visual Delay (sec) | Haptic Delay (sec) |
|---|---|---|
| Control | 0 | 0 |
| Anchoring | 0.25 | 0 |
| Anchoring | 0.5 | 0 |
| Anchoring | 0.75 | 0 |
| Synchronous | 0.25 | 0.25 |
| Synchronous | 0.5 | 0.5 |
| Synchronous | 0.75 | 0.75 |
| Asynchronous | 0.25 | 0.25 |
| Asynchronous | 0.5 | 0.25 |
| Asynchronous | 0.75 | 0.25 |

*Experiment Platform*

Building upon our detailed system design presented in Du et al. (2023), this section offers a concise overview of the key components of our teleoperation system, focusing on the VR system, its integration with various elements, and the implementation of delay coding functions.

Central to our teleoperation system is an advanced Virtual Reality (VR) setup, providing a fully immersive simulation environment developed in Unity. This platform replicates the physical dynamics and robot interactions with high fidelity, ensuring a realistic teleoperation experience. Another critical element in our system is the seamless integration between the Robot Operating System (ROS) and the Unity-based VR environment, facilitated by ROS#. This connection allows for real-time synchronization between the virtual environment and the physical robot, ensuring that any action taken in the VR space is instantly mirrored in the robot's movements.

To enhance the realism and interactivity of the VR environment, we incorporated the Touch X haptic controller. This device provides haptic feedback, replicating the physical sensations of manipulating objects or encountering resistance, crucial for tasks requiring fine motor control. The haptic feedback system is intricately coded to respond to both the operator's actions and the simulated physics of the VR environment, creating a cohesive and immersive experience. Finally, recognizing the impact of feedback delays on teleoperation, our system architecture includes specially developed coding functions to simulate various delay scenarios. Both visual and haptic feedback can be intentionally delayed, allowing us to study the operator's adaptability and performance under different sensory delay conditions.

*Data Collection Methods*

Optimal data collection quality for fNIRS requires careful preparation. Participants were advised to ensure their hair was clean and free from products that could obstruct the fNIRS sensors, and to avoid hairstyles or accessories that might disrupt the cap's placement. This preparation stage was critical for enhancing sensor-skin contact and the fidelity of the collected data, enabling a more accurate assessment of the cortical activity associated with the cognitive demands of the task. The stability of the experimental conditions, including controlled lighting and the participant's stationary posture while operating the haptic device, ensured that data integrity was maintained.

In the beginning, participants were asked to sign an informed consent form and fill out a background questionnaire about their age, gender, and VR experience. The experimental scene and content of each trial were the same. The sequence of tasks under different conditions was shuffled to eliminate the learning effects. The training session was designed to familiarize participants with the VR system and interactions within the virtual environment. Each participant was instructed to be acquainted with the devices (VR headset and haptic controller) and the virtual environment. Then, participants were given instructions about how to use the haptic controller to pick up and place the objects. After the training session, participants were asked to perform the pick-up and place task based on the virtual pipe skid system. After each trial, participants provided feedback through NASA TLX questionnaires.

During the experiment, participants were required to precisely control the robot gripper to stably grasp the cubes without knocking them away. Once successfully grasped, they should control the robot gripper past the obstacles and accurately place them on the corresponding target plate. The accuracy of the cube's positioning on the target is crucial, as it is a key metric for evaluating participants' operational performance. The use of visual and haptic feedback delays in the experimental design was critical for simulating the temporal challenges inherent in teleoperation tasks. These delays required the participants to rely on their cognitive adaptability, a phenomenon that conventional behavioral metric might not fully capture. This is the importance of fNIRS in our study. By employing fNIRS, we aimed to understand the fundamental neurological mechanisms, specifically, the impact of delays on the brain's activity during task performance and clarify the insights they provide into the cognitive strategy employed by the operator. This technique provided the method to measure the operator's brain activities in response to sensory feedback delays, offering objective data on the neural correlates of delay adaptation in teleoperation.

**RESULTS**

*Participants*

We recruited a total of 41 subjects for this experiment. The demographic information includes the gender, age group, major, and VR experience of participants are illustrated in **Table 2**. All participants reported that they were right-handed and did not have any known motor disorders or a history of neurological abnormalities.

**Table 2.** Demographic information of the participants

|  |  | Number | Percentage |
|---|---|---|---|
| Gender | Male | 26 | 63.41% |
|  | Female | 15 | 36.59% |
| Age Group | 18-24 | 14 | 34.15% |
|  | 25-30 | 24 | 58.53% |
|  | 31 and older | 3 | 7.31% |
| Major | Engineering (Civil, Coastal, Construction, Mechanical and related) | 18 | 43.90% |
|  | Non-Engineering | 23 | 56.10% |
| VR Experience | Experience with VR | 12 | 29.27% |
|  | Non-Experience with VR | 29 | 70.73% |

*Performance Results*

In our previous study Du et al. (2023), we investigated various performance metrics to determine the influence of delayed feedback in teleoperation. The placement accuracy, time on task, and cognitive load during pick-up and drop-off phases were evaluated using pupil size as a physiological indicator. Subjective assessments were also employed through the NASA Task Load

Index (NASA-TLX) questionnaire to measure the perceived workload and stress levels of participants.

For placement accuracy, we measured it as the Euclidean distance between the actual placement of the cube and the center of the target location. As illustrated in **Fig.4**, the results indicate that for the placement accuracy, the control condition is significantly better than asynchronous condition ($p=0.007$) as well as synchronous condition ($p=0.004$); the anchoring condition is significantly better than the asynchronous condition ($p=0.043$) and the synchronous condition ($p=0.032$). There is no significant difference between the control and anchoring ($p=0.168$), the asynchronous and synchronous condition ($p=0.892$). Time on Task is the difference between the end time and the start time of the task. The results also indicate significant differences between the control and anchoring condition ($p=0.009$), asynchronous condition ($p<0.001$), synchronous condition ($p<0.001$); and between anchoring and asynchronous condition ($p=0.018$) as well as the synchronous condition ($p=0.049$). There is no significant difference between the asynchronous and synchronous condition ($p=0.741$).

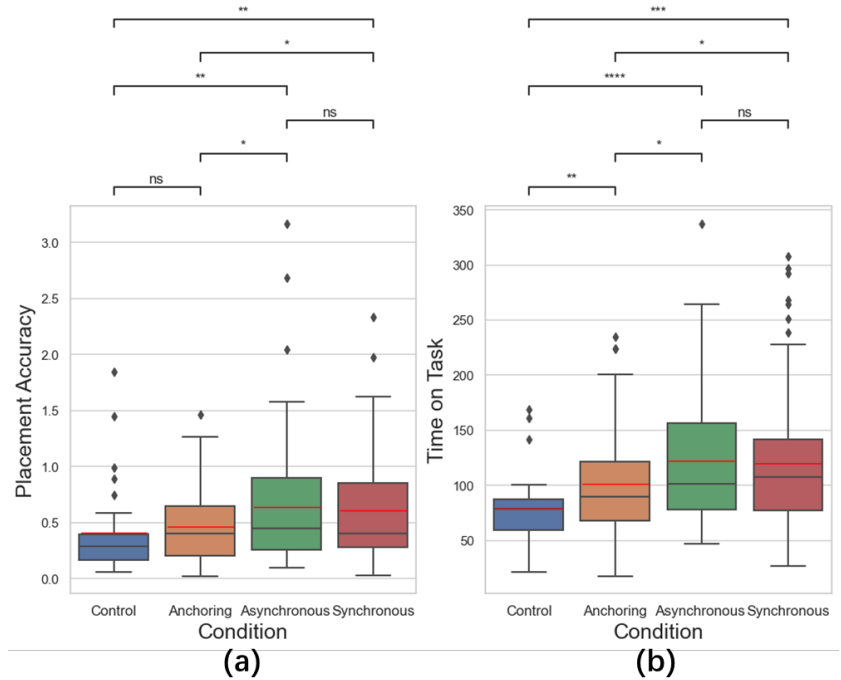

**Fig.4** Teleoperation (a) placement accuracy and (b) time on task comparison

About time perception, we focused on examining visual perception difference: the difference between the perceived visual delay ($Delay_{vp}$) and the actual visual delay ($Delay_{va}$); haptic perception difference: the difference between the perceived haptic delay ($Delay_{hp}$) and the actual haptic delay ($Delay_{ha}$); and visuomotor gap perception difference the difference between the perceived visuomotor gap ($Gap_p$) and the actual visuomotor gap ($Gap_a$). **Fig.5** shows the results of perception performance.

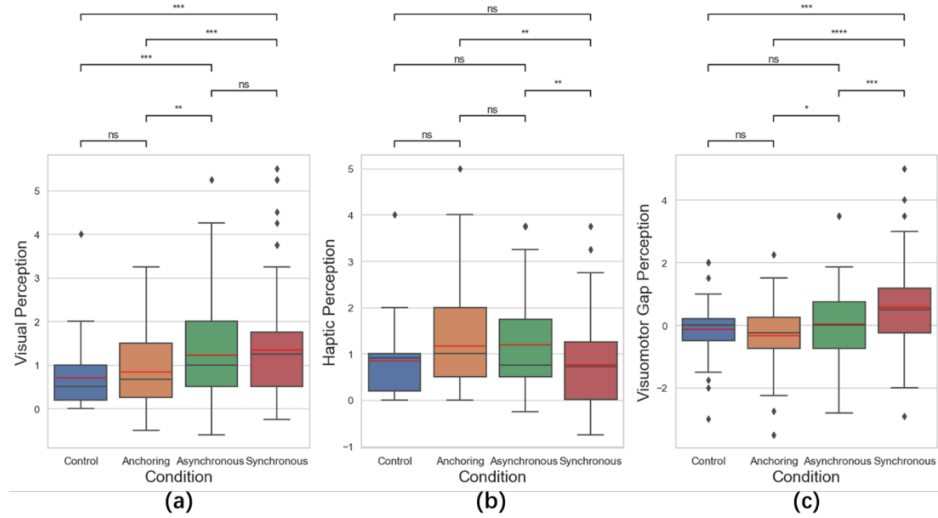

**Fig.5** Perception performance. (a) visual perception difference; (b) haptic perception difference; (c) visuomotor gap perception difference.

The results show that for visual perception difference, the control is significantly lower than asynchronous condition (p<0.001) and synchronous condition (p<0.001); anchoring condition is significantly lower than the asynchronous condition (p=0.003) as well as the synchronous condition (p<0.001). There is no significant difference between the control and anchoring (p=0.448), the asynchronous and synchronous condition (p=0.506). For the haptic perception difference, the results indicate that synchronous condition is significantly lower than anchoring condition (p=0.003) as well as asynchronous condition (p=0.001). There is no significant difference between the control and anchoring condition (p=0.091), asynchronous condition (p=0.090), synchronous condition (p=0.052); between anchoring and asynchronous condition (p=0.098). For the visuomotor gap perception difference, synchronous condition is significantly larger than control condition (p<0.001), anchoring condition (p<0.001), asynchronous condition (p<0.001); anchoring condition is lower than asynchronous condition (p=0.024). There is no significant between control and anchoring condition (p=0.237) and asynchronous condition (p=0.534).

For cognitive load, we developed a novel approach to evaluate participants' real-time cognitive load based on their pupillary diameter data (mm) collected by eye trackers. We divided the data of each trail into object pick-up stage and object drop-off stage. As illustrated in **Fig.6**, for the pick-up stage, the control condition is superior to anchoring condition (p=0.032), asynchronous condition (p=0.003), synchronous condition (p<0.001); anchoring condition have lower cognitive load than synchronous condition (p=0.004). There is no significant difference between anchoring and asynchronous condition (p=0.086), between asynchronous and asynchronous condition (p=0.276). For the drop-off stage, the results indicate that the control also better than asynchronous condition (p=0.006) and synchronous condition (p=0.012); the anchoring condition superior to asynchronous condition (p=0.048) and the synchronous condition (p=0.045). There is no significant difference between the control and anchoring condition (p=0.178), between the asynchronous and synchronous condition (p=0.983).

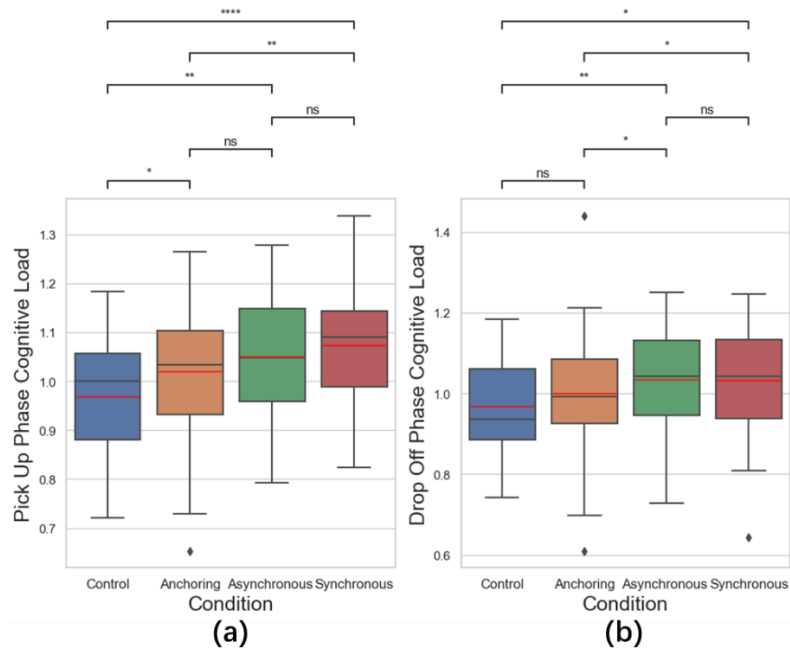

**Fig.6** Cognitive load changes in (a) object pickup and (b) drop-off stages.

The NASA-TLX results shown in **Fig.7**. The results indicate that for total score, control condition have the lowest cognitive load compared to anchoring condition (p=0.021), asynchronous condition (p=0.006), synchronous condition (p=0.024); There is no significant difference between anchoring and asynchronous condition (p=0.470) as well as the synchronous condition (p=0.843); between the asynchronous and synchronous condition (p=0.632). For confidence level, control condition also shows highest confidence level compared to anchoring condition (p=0.007), asynchronous condition (p<0.001), synchronous condition (p<0.001); anchoring condition is significantly higher than asynchronous condition (p=0.024) as well as the synchronous condition (p=0.019). There is no significant difference between the asynchronous and synchronous condition (p=0.829). For frustration level, control condition still superior to anchoring condition (p=0.004), asynchronous condition (p<0.001), synchronous condition (p<0.001); anchoring shows lower frustration level than asynchronous condition (p=0.033). There is no significant difference between anchoring and synchronous condition (p=0.110), between asynchronous and synchronous condition (p=0.694).

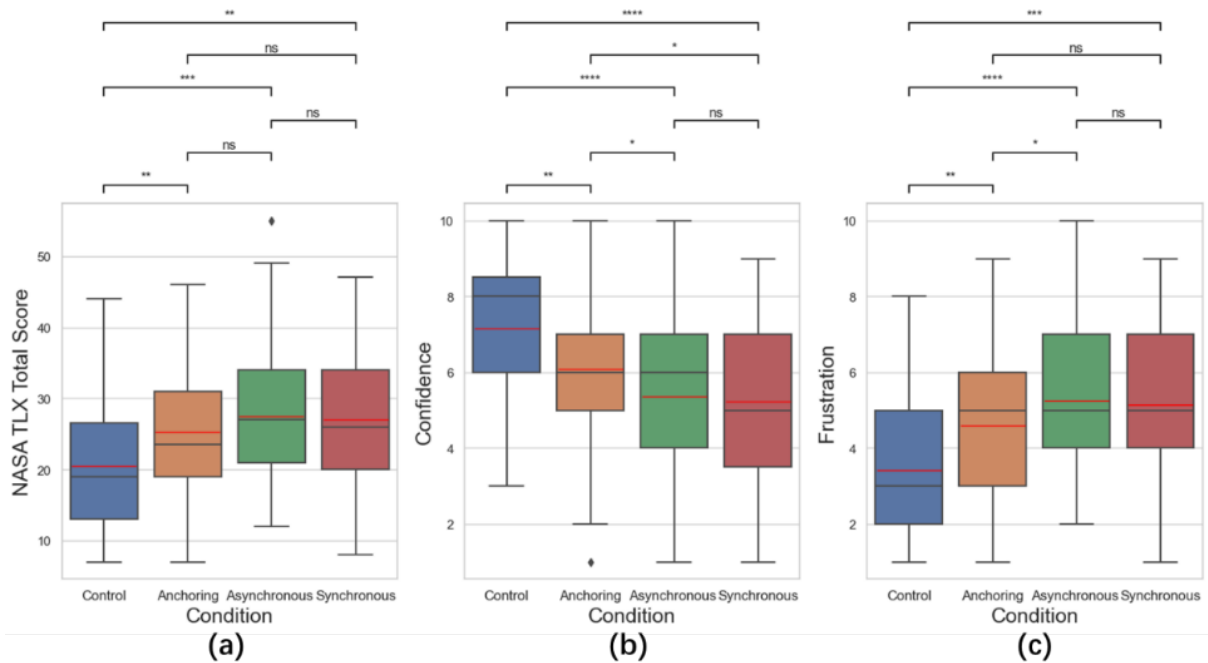

**Fig.7** NASA TLX result related to (a) total score (when calculating the total score, $10 - Confidence\ score$ is used as the calculation parameter), (b) self-confidence level, and (c) frustration level for delays up to 1s. Other NASA TLX results are not shown because of the insignificant difference among the conditions.

participants in the control condition reported higher levels of self-confidence and lower levels of frustration compared to other conditions, with anchoring also outperformed the synchronous and asynchronous conditions. The results from these metrics provided an initial understanding of the operational performance and cognitive states of operators under different feedback conditions.

Building upon this foundation, the present study delves deeper into the cognitive activities in different brain areas. By using fNIRS, we aim to demonstrate the specific brain regions engaged during teleoperation tasks, thereby providing a more refined perspective on the neural correlates of performance and brain activation. This approach allows us to pinpoint the hemodynamic responses in areas critical for decision-making, sensorimotor coordination, and time perception, factors that are critical to managing the challenges posed by feedback delays in teleoperation.

*fNIRS Results*

After the initial processing of the collected data, the raw intensity values captured by the NIRx fNIRS device were converted into optical density (OD) measurements. We then applied a combination of the Scalp Coupling Index (SCI) and a band-pass filter to process the OD data further. **Fig. 8 (a)** illustrated the raw OD data as initially recorded during the teleoperation tasks and **Fig. 8 (b)** illustrated the filtered OD data. The SCI was used to identify and exclude channels with insufficient signal quality, which show as lighter lines in filtered data. The remaining channels were then subjected to a bandpass filter, carefully designed to remove physiological noise such as cardiac and respiratory influences while preserving the signals pertinent to cognitive activity. These filtered OD values were then further processed to derive the concentration changes of oxyhemoglobin based on Beer-Lambert Law.

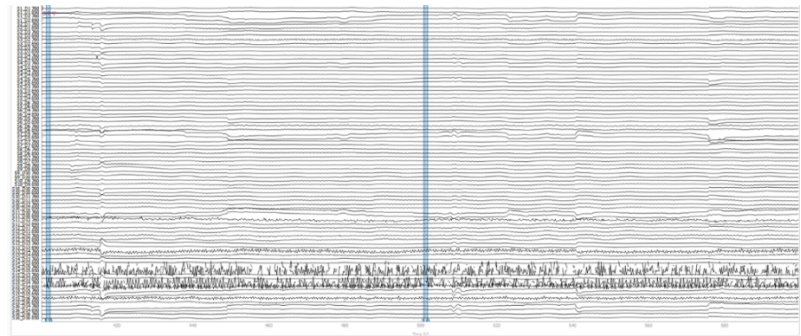
(a) Raw optical density signals

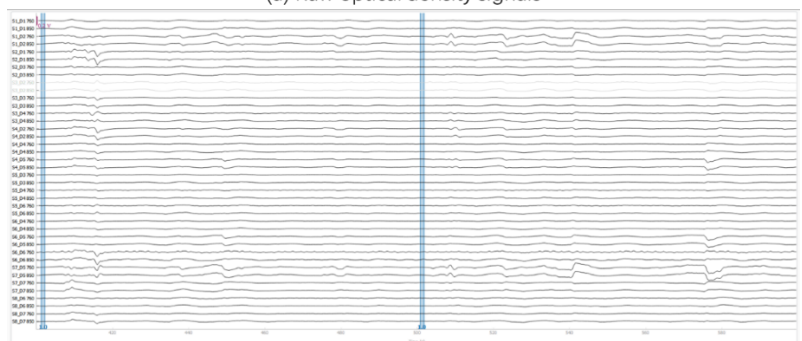
(b) Filtered optical density signals

**Fig.8** fNIRS signal preprocessing example of Participant #11: (a) raw optical density signals; (b) filtered optical density signals.

To analyze the brain activities to task events in teleoperation, we segmented the processed fNIRS data into specific epochs. Each epoch ranges from 10 seconds before to 30 seconds after the events of object pick-up and drop-off. **Fig.9** presents an example of this segmentation, showcasing data from participant #11 during a pick-up event. The figure visualizes the changes in oxyhemoglobin concentration, reflecting the brain's hemodynamic response during this critical phase of the task.

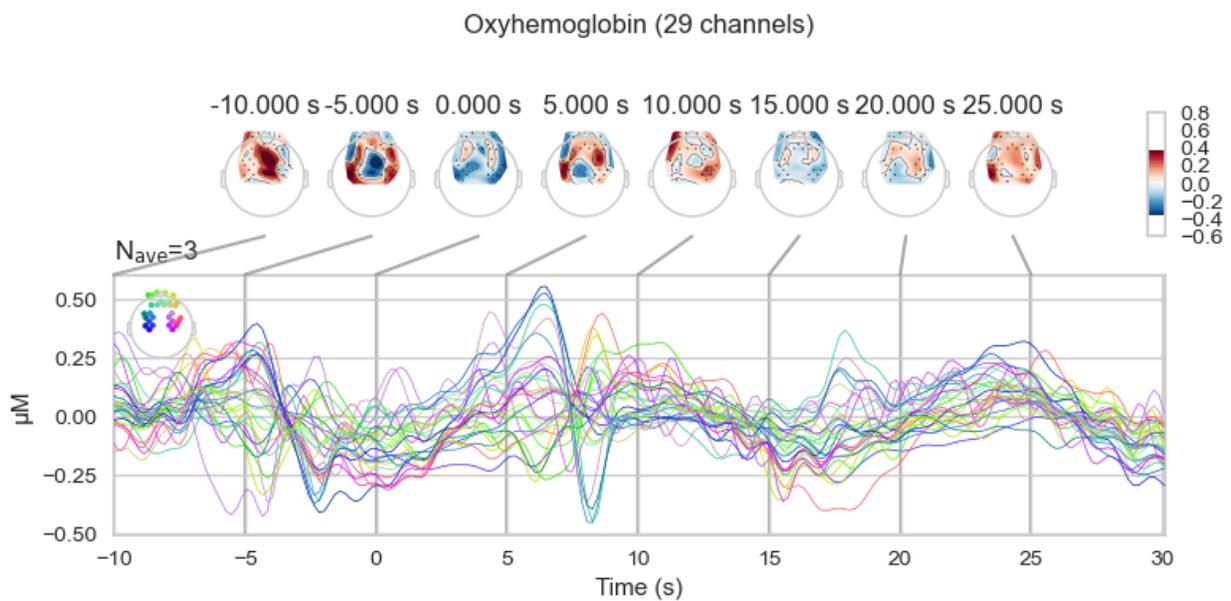

**Fig.9** Oxyhemoglobin concentration changes of the pick-up event for participant #11

To comprehensively evaluate the impact of different teleoperation conditions on brain activity, we conducted a statistical analysis of the oxyhemoglobin concentration across various brain areas. We employed the Kruskal-Wallis test, a non-parametric method used to determine if there are statistically significant differences between the groups. It is especially useful when our data does not follow a normal distribution, which is often the case in real-world data. The test essentially assesses whether one group is stochastically larger than the other and provides a p-value that we can use to test our hypothesis.

*Anterior Prefrontal Cortex Results*

As illustrated in **Fig.10**, in the anterior prefrontal cortex, known for its role in executive functions and decision-making, the anchoring condition showed lower brain activation compared to the asynchronous (p=0.005) and synchronous conditions (p=0.006). There is no significant difference between control and anchoring conditions (p=0.126), asynchronous conditions (p=0.062), synchronous conditions (p=0.063); between asynchronous and synchronous conditions (p=0.883). This could suggest that immediate haptic feedback, even when visual feedback is delayed, may help reduce the cognitive demands associated with integrating sensory information and making decisions. This reduction in brain activation could facilitate more efficient task performance, as the operator may rely more on the sense of touch, which is less affected by the delays.

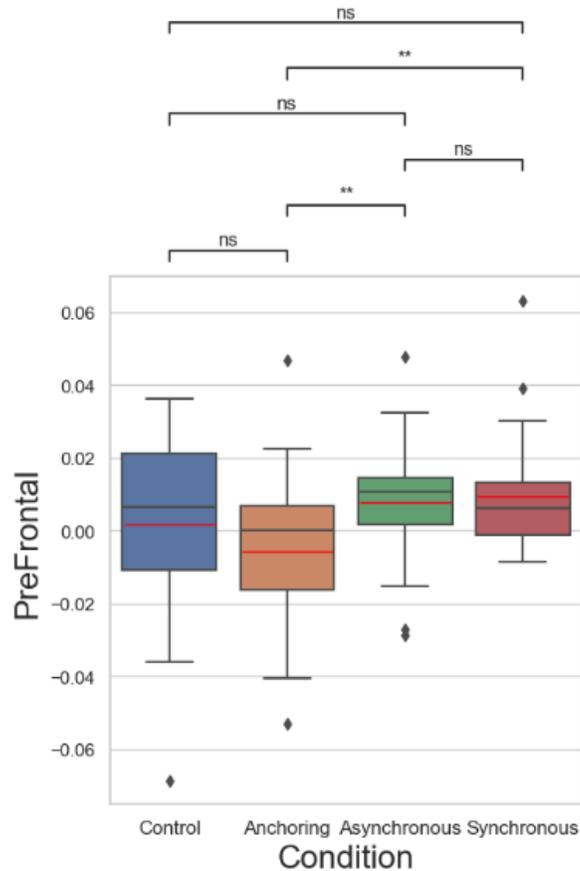

**Fig.10** Statistical analysis results of oxyhemoglobin concentrations changes in anterior prefrontal cortex

*Dorsolateral Prefrontal Cortex Results*

As illustrated in **Fig.11**, in the dorsolateral prefrontal cortex, associated with motor planning, working memory, and the cognitive aspects of time perception, exhibited a pattern of reduced brain activation in the anchoring condition. The left dorsolateral prefrontal cortex displayed a lower brain activation in both the control (p=0.006) and anchoring (p=0.017) conditions than in the synchronous condition. There is no significant difference between control and anchoring conditions (p=0.993) as well as asynchronous conditions (p=0.073); between anchoring and asynchronous conditions (p=0.095); and between asynchronous and synchronous conditions (p=0.392). The right dorsolateral prefrontal cortex exhibited a lower brain activation in the anchoring condition compared to both the asynchronous (p=0.002) and synchronous (p=0.003) conditions. There is no significant difference between control and anchoring conditions (p=0.113), asynchronous conditions (p=0.551), synchronous conditions (p=0.462); between asynchronous and synchronous conditions (p=0.749). This observation suggests that synchronized delays in feedback may hinder the operators' ability to effectively plan motor actions and manage time-based decision-making, consequently increasing brain activation. The anchoring condition, which provided immediate haptic feedback, appeared to promote a more efficient cognitive process, possibly by aiding in the temporal synchronization of motor actions and mitigating the disorienting effects of delayed visual feedback. It also highlights how the integration of haptic cues can support the cognitive processes involved in time perception, helping operators to maintain a coherent sense of timing despite the inherent delays in teleoperation.

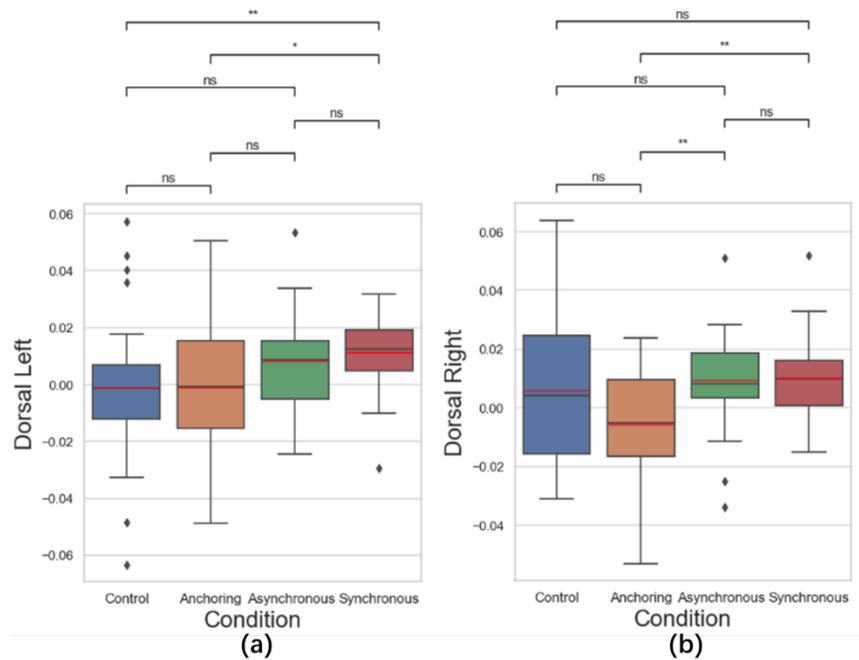

**Fig.11** Statistical analysis results of oxyhemoglobin concentrations changes in (a) left dorsolateral prefrontal cortex and (b) right dorsolateral prefrontal cortex

*Primary Motor Cortex Results*

As illustrated in **Fig.12**, in the primary motor cortex, responsible for the execution of movements, anchoring condition demonstrated better performance compared to the asynchronous condition. The left primary motor cortex displayed a lower brain activation in the anchoring condition compared to the asynchronous condition ($p=0.037$). There is no significant difference between control and anchoring conditions ($p=0.539$), asynchronous conditions ($p=0.180$), synchronous conditions ($p=0.993$); between anchoring condition and synchronous condition ($p=0.312$); between asynchronous and synchronous conditions ($p=0.113$). The right primary motor cortex also exhibited a lower brain activation in the anchoring condition compared to the asynchronous condition ($p=0.040$). There is no significant difference between control and anchoring conditions ($p=0.952$), asynchronous conditions ($p=0.243$), synchronous conditions ($p=0.517$); between anchoring condition and synchronous condition ($p=0.204$); between asynchronous and synchronous conditions ($p=0.243$). This suggests that the stabilizing effect of immediate haptic feedback extends beyond planning and preparation, directly facilitating the actual motor execution. The reduction in brain activation observed in this region further supports the idea that the immediate feedback in the anchoring condition mitigates the challenges brought on by delayed visual feedback, enhancing motor execution efficiency.

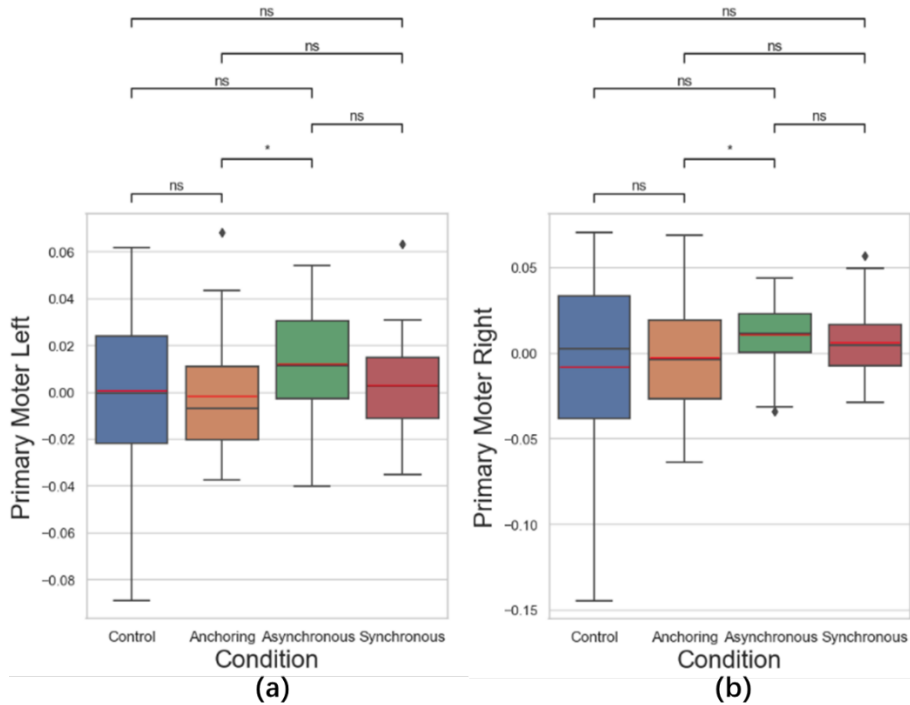

**Fig.12** Statistical analysis results of oxyhemoglobin concentrations changes in (a) left primary motor cortex and (b) right primary motor cortex

*Premotor Cortex Results*

As illustrated in **Fig.13**, in the premotor cortex, focused on the organization and planning of movements, anchoring condition also showed better performance compared to the asynchronous condition. The left premotor cortex displayed a lower brain activation in the anchoring condition compared to the asynchronous condition ($p=0.039$), suggesting that the immediate haptic feedback provided by the anchoring condition enhances the brain's ability to plan and prepare for movements. There is no significant difference between control and anchoring conditions ($p=0.431$), asynchronous conditions ($p=0.198$), synchronous conditions ($p=0.462$); between anchoring condition and synchronous condition ($p=0.058$); between asynchronous and synchronous conditions ($p=0.550$). For right premotor cortex, there is no significant difference between control and anchoring conditions (p=0.723), asynchronous conditions (p=0.076), synchronous conditions (p=0.634); between anchoring condition and asynchronous condition (p=0.186) as well as synchronous condition (p=0.452); between asynchronous and synchronous conditions (p=0.257). This finding indicates that even in the presence of visual feedback delays, immediate haptic feedback can effectively support the cognitive processes involved in organizing motor actions, leading to more efficient motor planning and reduced brain activation.

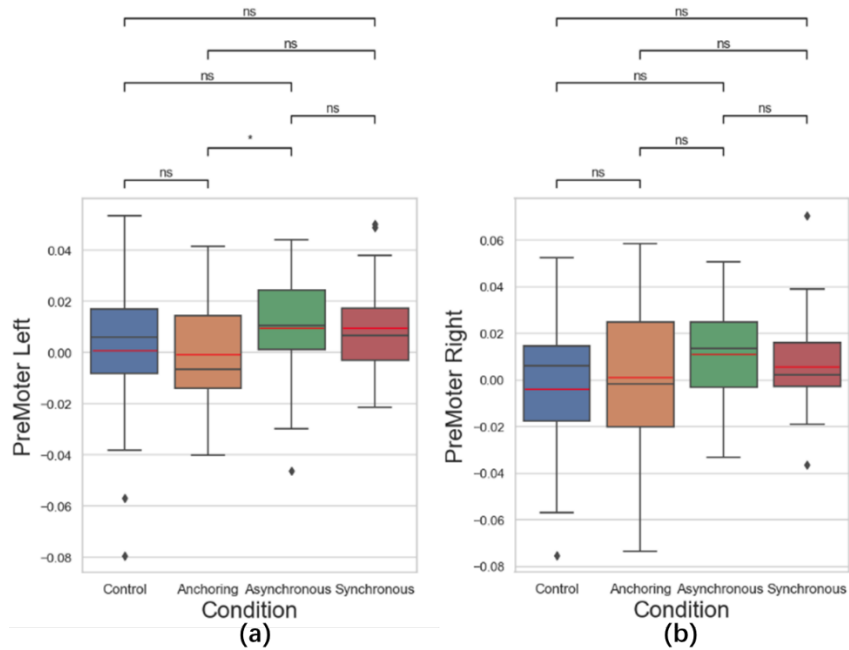

**Fig.13** Statistical analysis results of oxyhemoglobin concentrations changes in (a) left premotor cortex and (b) right premotor cortex

Interestingly, despite the control condition demonstrating better task performance, it was associated with a larger brain activity mean value across several cortical areas, including the prefrontal, right dorsolateral prefrontal cortex, and motor cortices. One possible explanation for this phenomenon is that: in the control condition, without feedback delay, operators may adopt a strategy that emphasizes speed and accuracy, taking advantage of the immediacy of the system's responses. This could result in the utilization of more "cognitive energy" to maintain a high level of performance. The term "cognitive energy" here refers to the engagement and allocation of cognitive resources, such as attention, working memory, and executive functions, that are necessary to perform a task effectively.

Consequently, the fNIRS data indicated increased activity in the relevant brain regions, which might reflect this intensive cognitive engagement. This high level of activation could be interpreted positively as an indicator of the operators' active and focused state, enabling them to perform efficiently without delays. Conversely, in the anchoring condition and even more so in the asynchronous and synchronous conditions, the presence of feedback delays may require a shift in strategy. Operators had to first compensate for the "disruption" introduced by the delay, which could involve a more cautious approach, increased error-checking, or a reliance on alternative sensory feedback (like haptic cues in the anchoring condition). This shift could lead to a different pattern of brain activation, possibly a less intense one, as operators may spread their cognitive resources over a longer period due to the delay in feedback.

Therefore, the reduced activation in the anchoring condition compared to the control condition might be due to a more distributed brain activation over time, rather than a concentrated burst of cognitive activity to immediately respond to feedback. This interpretation suggests that the high activation in the control condition aimed at optimizing performance, whereas in the delayed conditions, cognitive efforts might be partly directed towards mitigating the negative impacts of delay.

It's important to note that these assumptions about the nature of cognitive activation are based on the observed data patterns and theoretical understanding of task demands. However, without direct evidence of the operators' strategies or subjective experiences, these interpretations remain speculative. Further research, perhaps incorporating qualitative data on operator strategies or additional quantitative measures, would be necessary to substantiate these hypotheses.

**DISCUSSION**

Our human-subject experiment, designed to understand the neurofunctional implications of sensory manipulation in delayed robot teleoperation, yielded several insightful findings. Initially, when considering the neural data averaged across all phases of the experiment (pick-up, movement, and drop-off), no significant differences were observed among the four conditions: control, anchoring, synchronous, and asynchronous. Nevertheless, a focused analysis on the pick-up phase (40s) indicated differences among the four conditions. It suggests that the neurofunctional changes may have been event driven. And the pick-up phase represented a more difficult motor action, because the participants needed to move the robotic gripper to the center of the object, align well with the edge, and then grab the object, it did require more nuanced controls. While in contrast, the movement and the drop-off of the object on the target platform were comparably easier. As a result, we focused on the analysis of the pick-up phase.

In this phase, the anchoring condition (immediate simulated haptic feedback with delayed visual cue) not only improved motor performance but also led to a lower activation level in the dorsolateral prefrontal cortex (DLPFC), aligning closely with the control condition. This suggests that real-time synthetic force feedback might alleviate the cognitive burden associated with time perception challenges, potentially leading to faster actions and shorter times on tasks. Contrastingly, the asynchronous condition, with misaligned visual and haptic delays, resulted in the highest DLPFC activation, indicating increased cognitive strain in processing these delays. Furthermore, the anchoring condition was observed to reduce activation in the motor cortex, intriguingly even lower than in the control condition and significantly lower than in the asynchronous condition. This reduction could indicate enhanced efficiency in motor coordination and activity planning under the anchoring condition. Similarly, activation in the prefrontal cortex was lower in the anchoring condition compared to both the control and asynchronous conditions. This finding may imply that providing a consistent haptic cue, despite a delayed visual one, allows participants to rely more heavily on haptic feedback, reducing overall brain activation. This contrasts with the control condition, where participants might alternate between visual and haptic cues, possibly increasing cognitive load. However, it is also conceivable that the higher activation levels in the control condition may not necessarily reflect a negative aspect but could indicate positive engagement in the task. This area warrants further investigation to discern whether increased activation correlates with enhanced task engagement or cognitive strain.

    The reduced activation in the motor cortex under the anchoring condition, even lower than in the control condition, may reflect a more streamlined and efficient motor control process. According to studies like (Fitts and Posner 1967), as motor skills become more automated, the reliance on cognitive processes decreases, leading to reduced cortical activation. In the anchoring condition, the immediate haptic feedback might facilitate quicker motor learning and automation, thereby reducing the need for active motor planning and decision-making processes, typically associated with higher cortical activation. This efficiency could be attributed to a form of 'sensorimotor tuning', where the brain quickly adapts to the reliable haptic cues, optimizing motor outputs with less cognitive intervention (Wolpert et al. 2011). In addition, the lower activation in the prefrontal cortex in the anchoring condition suggests a reduction in cognitive load. This aligns with the theory of cognitive load proposed by (Sweller 1988), which posits that tasks with lower intrinsic cognitive demand result in lower cortical activation. By providing consistent haptic feedback, the anchoring condition may streamline the cognitive process, reducing the need for

continuous cross-modal integration and error-checking that is more pronounced in conditions with asynchronous or no feedback. This reduction in cross-modal processing, as discussed in the multisensory integration literature (Stein and Stanford 2008), may lead to a more efficient cognitive process with less prefrontal engagement.

The higher activation levels in the control condition present an intriguing paradox. One possibility, as suggested by studies in the field of cognitive neuroscience, is that this higher activation represents a positive engagement with the task (Jansma et al. 2000). Engaging actively with multiple sensory channels, as in the control condition, might stimulate more extensive neural networks, reflecting a more involved and perhaps even enjoyable task experience. However, this higher activation could also indicate a cognitive strain. The need to constantly switch between visual and haptic feedback, as theorized by (Alport et al. 1994), might place additional demands on cognitive resources, thereby increasing cortical activation. This scenario aligns with the dual-task interference model, which suggests that managing multiple streams of sensory information can elevate cognitive load (Pashler 1994). These observations underscore the complex interplay between sensory feedback, motor coordination, and cognitive processing in teleoperation. The anchoring condition, by providing immediate haptic feedback, seems to streamline both motor and cognitive processes, potentially offering a more efficient and less cognitively demanding approach to teleoperation. However, the higher activation in the control condition raises questions about the qualitative aspects of task engagement versus cognitive strain. Future research should aim to disentangle these aspects, possibly using subjective measures of task engagement and cognitive strain in conjunction with neuroimaging data. Additionally, exploring variations in task complexity and sensory feedback modalities could provide deeper insights into optimizing teleoperated systems for both performance efficiency and user experience.

**CONCLUSIONS**

This research is driven by the motivation to understand the neurofunctional implications of sensory manipulation in delayed robot teleoperation, a field that, despite its technological advancements, still hindered by the challenges of communication delays. The primary goal of this research is to fill a critical knowledge gap: the lack of neurofunctional evidence regarding the impact of simulated, synthetic haptic feedback on neural functions, especially those related to time perception and motor coordination. Delays in teleoperation can significantly affect performance, but the underlying neural dynamics, particularly in the context of sensory augmentation, remained largely unexplored. By focusing on these aspects, our study aims to provide insights that could lead to more intuitive and effective teleoperated systems, especially in applications demanding precision and timeliness.

Our human-subject experiment, involving different conditions of sensory feedback in teleoperation, revealed that the anchoring condition, with immediate simulated haptic feedback, not only improved motor performance but also regulated the activation levels of key brain regions such as the DLPFC and the motor cortex. This finding is significant as it suggests that providing real-time synthetic force feedback can reduce the cognitive and motor challenges posed by delayed teleoperation, particularly in the more demanding pick-up phase of the task. The reduction in DLPFC and motor cortex activation under the anchoring condition points towards a potential decrease in cognitive load and enhanced motor coordination. These results contribute to the understanding of how synthetic sensory feedback can be optimized to improve teleoperated task performance, providing a foundation for future technological developments in this area.

While our findings are promising, they are not without limitations. The study's scope was confined to a controlled experimental setting, which might not fully capture the complexities of real-world teleoperation scenarios. Additionally, the focus on specific brain regions, though insightful, does not encompass the entire spectrum of neural processes involved in teleoperation. Future research should aim to replicate these findings in more varied and dynamic settings to verify their applicability in real-world applications. Furthermore, exploring other forms of sensory manipulation and their neurofunctional impacts, as well as investigating the long-term effects of such interventions on skill acquisition and adaptation in teleoperation, would be beneficial. These future agenda items could provide deeper insights into the neural mechanisms underlying teleoperated systems, guiding the development of more responsive, efficient, and user-friendly teleoperation technologies.

## DATA AVAILABILITY

All data used in this paper can be found at: https://www.dropbox.com/scl/fo/4qcvc4v3nxhc9in0oi6bo/h?rlkey=15eds1e7on4vjkstannruok5r&dl=0

## ACKNOWLEDGEMENTS

This material is supported by the National Science Foundation (NSF) under grant 2024784 and The National Aeronautics and Space Administration (NASA) under grant 80NSSC21K0845. Any opinions, findings, conclusions, or recommendations expressed in this article are those of the authors and do not reflect the views of the NSF or NASA.